\documentclass[prd,twocolumn,showpacs,floatfix,amsmath,amssymb,floatfix]{revtex4}
\usepackage{graphicx,dcolumn,booktabs,bm}
\usepackage{longtable,lscape}
\usepackage{txfonts}
\usepackage{overpic}
\usepackage{amssymb}
\usepackage{indentfirst}
\usepackage{feynmf}   
\usepackage{slashed}  
\usepackage{cases}
\usepackage{color,ulem}

\graphicspath{{Figures/}} 

\begin{document}


\title{Charged charmoniumlike state $Z_c(3900)^\pm$ via meson photoproduction}
\author{Qing-Yong Lin$^{1,2,4}$}\email{qylin@impcas.ac.cn}
\author{Xiang Liu$^{3,4}$\footnote{corresponding author}}\email{xiangliu@lzu.edu.cn}
\author{Hu-Shan Xu$^{1,4}$}

\affiliation{
$^1$Institute of Modern Physics, Chinese Academy of Sciences, Lanzhou 730000, China\\
$^2$University of Chinese Academy of Sciences, Beijing, 100049, China\\
$^3$School of Physical Science and Technology, Lanzhou University, Lanzhou 730000, China\\
$^4$Research Center for Hadron and CSR Physics, Lanzhou University and Institute of Modern Physics of CAS, Lanzhou 730000, China}

\date{\today}

\begin{abstract}
In this work, we explore the production of the newly observed charged charmoniumlike state $Z_c(3900)^\pm$ by the photoproduction process $\gamma p\to Z_c(3900)^+n$. Our numerical result indicates that the maximum of the calculated total cross section of $\gamma p\to Z_c(3900)^+n$
can reach up to the order of $0.1 \mu b$. Furthermore, the Dalitz plot analysis is performed by considering the Pomeron exchange as the background contribution. This analysis shows that the $Z_c(3900)^+$ signal can be distinguished from the
background easily and the best energy window  of searching for $Z_c(3900)^+$
is $\sqrt{s}\simeq7$ GeV, which is important information for further experimental study of $Z_c(3900)^+$
by meson photoproduction.
\end{abstract}

\maketitle

\section{Introduction}\label{sec:intro}

Very recently, the BESIII Collaboration announced the observation of a
charged charmoniumlike state $Z_c(3900)^\pm$ in the $J/\psi\pi^{\pm}$ invariant mass spectrum of $e^+e^-\to J/\psi\pi^+\pi^-$at $\sqrt{s}=4.26$ GeV \cite{Ablikim:2013mio}. Later, this novel phenomenon was
confirmed by the Belle Collaboration in the same process \cite{Liu:2013dau} and by
Ref. \cite{Xiao:2013iha} in $e^+e^-\to J/\psi\pi^+\pi^-$ at $\sqrt{s}=4.17$ GeV. In addition, the corresponding neutral partner of $Z_c(3900)^\pm$ was also reported at a 3$\sigma$ level of significance \cite{Xiao:2013iha}. The observation of $Z_c(3900)^\pm$ makes the family of charmoniumlike state $X,Y,Z$ become abundant.
The experimental information of $Z_c(3900)^\pm$
from different experiments is listed in Table \ref{mass-width}. A peculiarity of $Z_c(3900)^\pm$ is that its mass is close to the $D\bar{D}^*$ threshold.

\begin{table}[htb]
\caption{The mass and width of $Z_c(3900)^\pm$ measured by BESIII, Belle, and given in Ref. \cite{Xiao:2013iha}.
\label{mass-width}}
\begin{tabular}{ccc}
\toprule[1pt]
Experiment & Mass (MeV) & Width (MeV)\\
\midrule[1pt]
BESIII \cite{Ablikim:2013mio} & $3899\pm3.6(\mathrm{stat})\pm 4.9(\mathrm{syst})$
& $46\pm 10(\mathrm{stat})\pm 20(\mathrm{syst})$ \\
Belle \cite{Liu:2013dau} & $3894.5\pm6.6(\mathrm{stat})\pm 4.5(\mathrm{syst})$
& $63\pm24(\mathrm{stat})\pm 26(\mathrm{syst})$ \\
Ref. \cite{Xiao:2013iha} & $3885\pm5(\mathrm{stat})\pm 1(\mathrm{syst})$
& $34\pm12(\mathrm{stat})\pm 4(\mathrm{syst})$ \\
\bottomrule[1pt]
\end{tabular}
\end{table}

Before the observation of $Z_c(3900)^\pm$, there are two charged bottomoniumlike states $Z_b(10610)$ and $Z_b(10650)$ reported by Belle \cite{Belle:2011aa}, which are near the $B\bar{B}^\ast$ and $B^\ast\bar{B}^\ast$ thresholds, respectively. Different theoretical explanations to $Z_b(10610)$ and $Z_b(10650)$ were proposed \cite{Bondar:2011ev,Chen:2011zv,Dias:2011mi,Bugg:2011jr,Zhang:2011jja,Yang:2011rp,Guo:2011gu,Sun:2011uh,Sun:2012zzd,Chen:2011pv,Ali:2011ug}.
It should be noted that the charmoniumlike structures near the $D\bar{D}^*$ and $D^*\bar{D}^*$ thresholds were predicted in Refs. \cite{Sun:2011uh,Sun:2012zzd,Chen:2011xk}, which can be as the partners of the observed $Z_b(10610)$ and $Z_b(10650)$. The $Z_c(3900)^\pm$ observation provides a good test to former theoretical predictions \cite{Sun:2011uh,Sun:2012zzd,Chen:2011xk}.

$Z_c(3900)^\pm$ has also inspired extensive studies on its underlying properties, where $Z_c(3900)^\pm$ was explained as molecular state
{\cite{Wang:2013cya,Cui:2013yva,Zhang:2013aoa,Ke:2013gia,Dong:2013iqa,Wilbring:2013cha}}, tetraquark
state \cite{Faccini:2013lda,Dias:2013xfa,Braaten:2013boa,Qiao:2013raa}, cusp effect \cite{Liu:2013vfa},
initial-single-pion-emission mechanism \cite{Chen:2013coa}, and so on. More discussions can be found in
Refs. \cite{Voloshin:2013dpa,Mahajan:2013qja}. At present, carrying out the investigation of $Z_c(3900)^\pm$ is an interesting research topic.

Besides the analysis of the mass spectrum and the decay behavior of $Z_c(3900)^\pm$, studying the production of $Z_c(3900)^\pm$ can provide important information on $Z_c(3900)^\pm$. However, the corresponding study has been absent since $Z_c(3900)^\pm$ was reported. Thus, in this work we focus on this problem and explore whether $Z_c(3900)^\pm$ can be produced by different processes from the $e^+e^-$ collision. In fact, for charmoniumlike states $Z(4430)^\pm$ \cite{Choi:2007wga} and $Y(3940)$ \cite{Abe:2004zs} observed in the $B$ meson decays, similar studies were given in Refs. \cite{Liu:2008qx,Ke:2008kf,He:2009yda}. Liu, Zhao and Close proposed that
$Z(4430)^\pm$ can be produced by the meson photoproduction process $\gamma p\to Z(4430)^+ n\to \psi^\prime \pi^+ n$ \cite{Liu:2008qx}. Later, the authors in Ref. \cite{Ke:2008kf} suggested to search for $Z(4430)^\pm$ by the nucleon-antinucleon scattering at the forthcoming PANDA experiment. The discovery potential for charmoniumlike state $Y(3940)$ by the meson photoproduction process $\gamma p\to Y(3940)p$ was discussed in Ref. \cite{He:2009yda}.

In this work, we study the production of $Z_c(3900)^\pm$ through the meson photoproduction. Our calculations cannot only reveal the discovery possibility of $Z_c(3900)^\pm$ by the meson photoproduction, but also provide crucial information of the suitable meson photoproduction process and the best energy window of searching for $Z_c(3900)^\pm$. Of course, it is valuable to further experimentally study $Z_c(3900)^\pm$.

This paper is organized as follows. After the introduction, we present
the calculation of the production of $Z_c(3900)^\pm$ via meson photoproduction. In Sec. \ref{sec:bg}, the
possible background contribution to the $J/\psi\pi^+n$ final state
is discussed, and the corresponding Dalitz plot analysis is given. This work ends with the discussion and conclusion.

\section{The $Z_c(3900)^\pm$ production in meson photoproduction}\label{sec:model}

\begin{figure}[htb]
\begin{center}
\scalebox{1.0}{\includegraphics[width=\columnwidth]{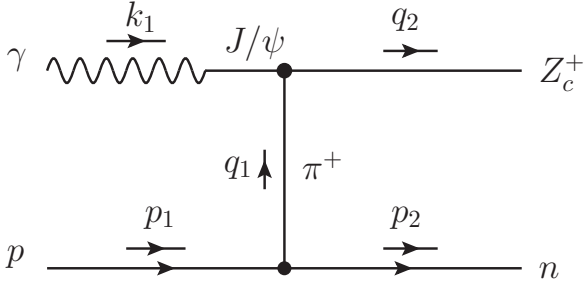}}
\caption{The $Z_c(3900)^+$ photoproduction through the pion exchange.
\label{fig:2to2}}
\end{center}
\end{figure}

We choose the photoproduction process $\gamma p \to Z_c(3900)^+ n$ to study the production of $Z_c(3930)^\pm$ since $Z_c(3900)^\pm$ can couple with $J/\psi\pi^\pm$ \cite{Ablikim:2013mio,Liu:2013dau}.
Due to the vector meson dominance (VMD) assumption \cite{Bauer:1975bv,Bauer:1975bw,Bauer:1977iq}, a photon can interact with $J/\psi$. And then $\gamma p\to Z_c(3900)^+n$ occurs via the pion exchange just shown in Fig. \ref{fig:2to2}.
First, we mainly concentrate on the production probability of $Z_c(3900)^+$ in the $\gamma p \to Z_c(3900)^+ n$ process, where the differential and total cross sections
of this photoproduction process are discussed.

Under the VMD model \cite{Bauer:1975bv,Bauer:1975bw,Bauer:1977iq},
the general expression of an intermediate vector meson $V$ coupling
with photon $\gamma$ is written as
\begin{equation}
{\cal L}_{V\gamma} = -\frac{eM_V^2}{f_V}V_\mu A^\mu,
\end{equation}
where $M_V$ and $f_V$ denote the mass and the decay constant of
the vector meson, respectively. The coupling constant $e/f_V$ can be determined by
the decay $V\to e^{+}e^{-}$, i.e.,
\begin{eqnarray}\label{eqa:vmd}
\frac{e}{f_V} &=& \left[\frac{3\Gamma_{V\to e^+ e^-}M_V^2}{8\alpha |\vec{p}|^3}\right]^{1/2} \simeq \left[\frac{3\Gamma_{V\to e^+ e^-}}{\alpha M_V}\right]^{1/2} \ ,
\end{eqnarray}
where $|\vec{p}|=(M_V^2-4m_e^2)^{1/2}/2\simeq M_V/2$ is the three momentum of an electron in the rest
frame of the vector meson. $\alpha = e^2/(4\pi) = 1/137$ means the fine-structure constant.
By $\Gamma_{J/\psi\to e^+e^-}=5.55\pm 0.14\pm 0.02$
keV~\cite{Beringer:1900zz} and Eq. (\ref{eqa:vmd}), we obtain $e/f_{J/\psi}= 0.027$.

For depicting the coupling between the pion and the nucleons, we use the
the effective Lagrangian
\begin{eqnarray}
{\cal L}_{\pi NN} &=& -\frac{g_{\pi NN}}{2m_N} \bar{N}\gamma_5\gamma_\mu
(\vec{\tau}\cdot\partial^\mu\vec{\pi}) N  \nonumber \\
&=& -\frac{g_{\pi NN}}{2m_N}(\bar{p}\gamma_5\gamma_\mu \partial^\mu\pi^0 p+\sqrt{2}\bar{p}\gamma_5\gamma_\mu
\partial^\mu\pi^{+} n \nonumber \\
&& + \sqrt{2}\bar{n}\gamma_5\gamma_\mu \partial^\mu\pi^{-}p - \bar{n}\gamma_5\gamma_\mu \partial^\mu\pi^0n),
\end{eqnarray}
where the coupling constant $g_{\pi NN}^2/4\pi=14$ is
adopted~\cite{Goldberger:1958tr}. In addition, a monopole form factor
\begin{equation}
F_{\pi NN} = \frac{\Lambda_\pi^2 - m_\pi^2}{\Lambda_\pi^2 - q^2}
\end{equation}
is introduced, where $q$ denotes the four momentum of the exchanged pion. The cutoff $\Lambda_\pi$ is a free parameter for the $\pi NN$ vertex. In the following, we will discuss the dependence of the results on the value of $\Lambda_\pi$.

If $Z_{c}(3900)^\pm$ is a resonance, $Z_{c}(3900)^\pm$ must be a good candidate of an exotic state since $Z_{c}(3900)^\pm$ is a charged charmoniumlike state \cite{Ablikim:2013mio,Liu:2013dau}. Among several possible exotic state assignments to $Z_{c}(3900)^\pm$, the S-wave $D\bar{D}^*$ molecular state is the most popular explanation \cite{Sun:2011uh,Sun:2012zzd}, where the spin-parity quantum number of $Z_c(3900)^\pm$ favors $J^P=1^{+}$. In the following, according to this quantum number assignment, we discuss the production of $Z_c(3900)^\pm$. The effective Lagrangian describing the
$Z_c(3900)^+J/\psi \pi$ coupling is {\cite{Liu:2008qx,Xiong:1992ui,Haglin:1994yv}}
\begin{eqnarray}
\mathcal {L}_{Z\psi\pi} &=& \frac{g_{Z\psi\pi}}{M_Z}\left( \partial^\mu \psi^{\nu} \partial_\mu \pi Z_\nu -
\partial^\mu \psi^{\nu} \partial_\nu \pi Z_\mu \right) ,
\end{eqnarray}
where the details of constructing the effective Lagrangian can be found in Ref. \cite{Xiong:1992ui}. In the following formula, we use $Z$ and $\psi$ to denote $Z_c(3900)$ and $J/\psi$, respectively.
The coupling constant $g_{Z\psi\pi}$ can be determined by
the decay width of $Z_c(3900)^+ \to \psi\pi^+$ \cite{Beringer:1900zz}, i.e.
\begin{eqnarray}\label{gamma-z}
\Gamma\left[Z^+ \to \psi \pi^+\right] &=& \frac{1}{8\pi}\frac{|{\vec{p}^\prime}|}{M_Z^2} \frac{|{\cal{M}}[Z^+ \to \psi \pi^+]|^2}{3} \nonumber\\
&=&\left(\frac{g_{Z\psi\pi}}{M_Z}\right)^2 \frac{ |{\vec{p}^\prime}|}{24\pi M_Z^2}
 \left[ (M_Z^2-M_\psi^2-m_\pi^2)^2/2 \right. \nonumber \\
 &&\left. + M_{\psi}^2(m_\pi^2+|{\vec{p}^\prime}|^2)\right] ,
\end{eqnarray}
where $\vec{p}^\prime=[(M_Z^2-(M_\psi+m_\pi)^2)(M_Z^2-(M_\psi-m_\pi)^2)]^{1/2}/(2M_Z)$ denotes the three momentum of the daughter mesons in the parent's center-of-mass frame. The mass of $Z_c(3900)^\pm$ is taken as $M_Z=3899$ MeV~\cite{Ablikim:2013mio}. Thus, we get $g_{Z\psi\pi}/M_Z=1.50$ and $g_{Z\psi\pi}/M_Z= 1.90$,
respectively, for typical values $\Gamma[Z^+\to\psi\pi^+]=29$ MeV from the estimate in Ref. \cite{Faccini:2013lda} and $\Gamma[Z^+\to\psi\pi^+]= 46$ MeV, which corresponds to the full width of $Z_c(3900)^\pm$ \cite{Ablikim:2013mio}.

For the vertex of $Z_c(3900)^+$ interacting with $J/\psi\pi^+$, we also introduce a form factor $F_{Z\psi\pi}$, which satisfies the form
\begin{equation}
F_{Z\psi\pi} = \frac{M_{\psi}^2 - m_i^2}{M_{\psi}^2 - q^2}
\end{equation}
with $m_i=m_\pi$. Here, the cutoff is set as the mass of the
intermediate vector meson~\cite{Friman:1995qm}, i.e., $\Lambda = M_{\psi}$.

With the above preparation, we finally obtain the amplitude of the $\gamma(k_1,\epsilon_\gamma) p(p_1) \to
Z_c(3900)^+(q_2,\epsilon_Z) n(p_2)$ process
\begin{eqnarray}
\mathcal{T}_{fi} &=& -i\left(\frac{g_{\pi NN}}{\sqrt{2}m_N}
\frac{g_{Z\psi\pi}}{M_Z} \frac{e}{f_{\psi}}\right)
\bar{u}(p_2)\gamma_5 \slashed{q}_1 u(p_1) \epsilon_Z^{*\mu} \epsilon_\gamma^\nu  \nonumber \\
&& \times [k_1\cdot(q_2-k_1)g_{\mu\nu}-k_{1\mu}(q_2-k_1)_\nu]  \nonumber \\
&& \times \frac{1}{q_1^2-m_\pi^2} F_{\pi NN}(q_1^2)
F_{Z\psi\pi}(q_1^2).
\end{eqnarray}
By defining $k \equiv (p_1 - p_2)^2 \equiv q_1^2$, $s\equiv
(k_1+p_1)^2$, the corresponding unpolarized differential cross section reads as
\begin{equation}
 \frac{d\sigma}{dk} = \frac{1}{64\pi s} \frac{1}{|\vec{k}_{1cm}|^2}
 \frac{|{\cal M}|^2}{4}
\end{equation}
with
\begin{eqnarray}\label{trans-v-axi}
|{\cal M}|^2 &=& \sum \left| \mathcal{T}_{fi}  \right|^2 \nonumber \\
&=&\left(\sqrt{2}g_{\pi NN} \frac{g_{Z\psi\pi }}{M_Z}
\frac{e}{f_{\psi}}\right)^2 \frac{-q_1^2(q_1^2-M_Z^2)^2}{(q_1^2-m_\pi^2)^2}   \nonumber \\
&& \times \left(\frac{\Lambda_\pi^2
-m_\pi^2}{\Lambda_\pi^2-q_1^2} \right)^2 \left(\frac{M_{\psi}^2
-m_\pi^2}{M_{\psi}^2-q_1^2} \right)^2.
\end{eqnarray}
The total cross section can be obtained by integrating over the range of $|k|$. 

\begin{figure}[htbp]
\begin{center}
\scalebox{0.95}{\includegraphics[width=\columnwidth]{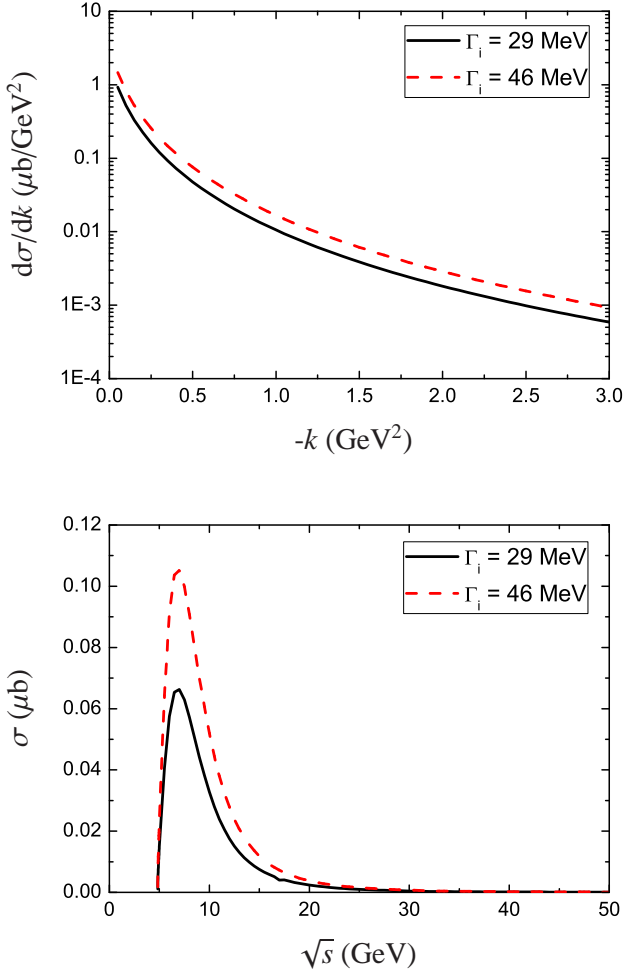}}
\caption{(color online). The obtained differential cross section (top) and total cross section (bottom) for $\gamma p \to Z_c\mbox{(3900)}^+n$. The black solid and red
dashed lines correspond to results when taking the typical values $\Gamma[Z^+\to\psi\pi^+]=29$ MeV and  $\Gamma[Z^+\to\psi\pi^+]=46$ MeV, respectively.
\label{fig:2to2res}}
\end{center}
\end{figure}

{In Fig.~\ref{fig:2to2res}, the
differential and total cross sections are plotted with $\Lambda_\pi =0.7$ GeV \cite{Liu:2008qx}}, where the differential cross section is given with the fixed $\sqrt{s} \simeq 7 \,\mbox{GeV}$. The differential cross section goes down with the increase of the $-k$ value. Since the total cross section is proportional to the square of the coupling $g_{Z\psi\pi}$, thus the total cross section is also proportional to the decay width of $Z_c(3900)^+\to J/\psi\pi^+$. At present, the experiment did not give the concrete measurement of this decay width, which makes us take two typical values of the decay width of $Z_c(3900)^+\to J/\psi\pi^+$ when discussing the differential and total cross sections. As discussed above, the theoretical value $\Gamma[Z^+\to\psi\pi^+]=29$ MeV was calculated in Ref. \cite{Faccini:2013lda}. Thus, the corresponding differential and total cross sections are listed in Fig. \ref{fig:2to2res}. If taking the full width of $Z_{c}(3900)^\pm$ as the input, we obtain the upper limit of differential and total cross sections of the $\gamma p \to Z_c\mbox{(3900)}^+n$ process.

\begin{figure}[htbp]
\begin{center}
\scalebox{0.95}{\includegraphics[width=\columnwidth]{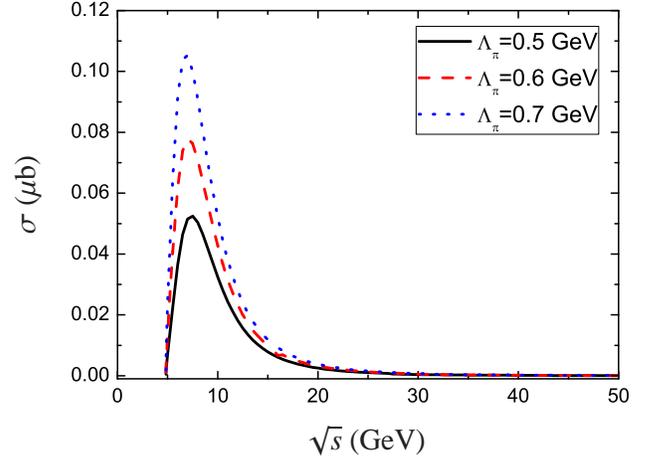}}
\caption{(color online). The obtained total cross section for $\gamma p \to Z_c\mbox{(3900)}^+n$ with the typical cutoff $\Lambda_\pi$. Here, the
partial decay width is taken as $\Gamma[Z^+\to\psi\pi^+]=46$ MeV.
\label{fig:2to2TCS-cutoff}}
\end{center}
\end{figure}

As shown in Fig. \ref{fig:2to2TCS-cutoff}, we also present the variation of the total cross sections to the
different typical $\Lambda_\pi$ values, where $\Lambda_\pi$ is taken as 0.5 GeV, 0.6 GeV and 0.7 GeV.
The total cross sections of $\gamma p \to Z_c\mbox{(3900)}^+n$ depend on the
value of $\Lambda_\pi$, i.e., the cross section changes by a factor of two when the cutoff is
varied from 0.5 GeV to 0.7 GeV.

\section{The background analysis}\label{sec:bg}

Besides giving the differential and total cross section of the $Z_{c}(3900)^\pm$ photoproduction, it is also important to further perform the background analysis of the corresponding reaction.
Thus, we consider the $\gamma p \to Z_c(3900)^+n \to J/\psi\pi^+n$ and $\gamma p \to J/\psi p \to J/\psi\pi^+n$ processes, which result in the signal and background contributions, respectively.
Here, $\gamma p \to J/\psi p \to J/\psi\pi^+n$ can occur via the Pomeron exchange \cite{Liu:2008qx}.
As a simple estimate of background, we only apply the Pomeron exchange model to illustrate the background contribution when exploring the discovery potential of $Z_c(3900)$ via meson photoproduction.
In the following, we will illustrate them systematically.

\subsection{The $\gamma p \to Z_c(3900)^+n \to J/\psi\pi^+n$ reaction}

\begin{figure}[htbp]
\begin{center}
\scalebox{0.9}{\includegraphics[width=\columnwidth]{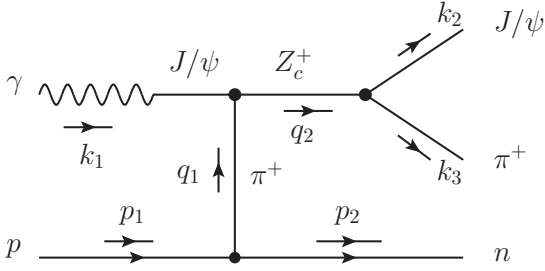}}
\caption{The diagram depicting $\gamma p\to Z_c(3900)^+n\to J/\psi\pi^+n$ through the $\pi^+$ exchange.
\label{fig:2to3s}}
\end{center}
\end{figure}

As the main signal contribution, the $\gamma p \to Z_c(3900)^+n \to J/\psi\pi^+n$ reaction is described in
Fig. \ref{fig:2to3s}.

The transition amplitude can be expressed as
\begin{eqnarray}\label{trans-v-axi}
\mathcal{T}_{fi}^{Singal} &=& -i\left(\frac{g_{\pi NN}}{\sqrt{2}m_N} \left(\frac{g_{Z\psi\pi }}{M_Z}\right)^2
\frac{e}{f_{\psi}}\right)\bar{u}(p_2)\gamma_5 \slashed{q}_1 u(p_1) \nonumber\\
&& \times\left(k_1\cdot q_1g^{\mu\beta}-k_1^\beta q_1^\mu\right)\left(k_2\cdot k_3g^{\alpha\nu}-k_2^\alpha k_3^\nu\right)\nonumber \\
&& \times \frac{1}{q_1^2-m_\pi^2}
\frac{-g_{\alpha\beta}+q_{2\alpha}q_{2\beta}/M_Z^2}{q_2^2-M_Z^2+i M_Z\Gamma_Z}\epsilon_{\gamma\mu}\epsilon_{\psi\nu}^\ast \nonumber \\
&& \times \left(\frac{\Lambda_\pi^2-m_\pi^2}{\Lambda_\pi^2-q_1^2} \right)\left(\frac{\Lambda_Z^2
-m_\pi^2}{\Lambda_Z^2-q_1^2} \right)\left(\frac{\Lambda_Z^2
-M_Z^2}{\Lambda_Z^2-q_2^2} \right),\label{h0}
\end{eqnarray}
where the cutoff is taken as $\Lambda_Z=M_\psi$ \cite{Friman:1995qm} and the relevant
kinematic variables are listed in Fig. \ref{fig:2to3s}.

\subsection{$\gamma p \to J/\psi\pi^+n$ with the Pomeron exchange}

\begin{figure}[htbp]
\begin{center}
\scalebox{0.4}{\includegraphics[width=\textwidth]{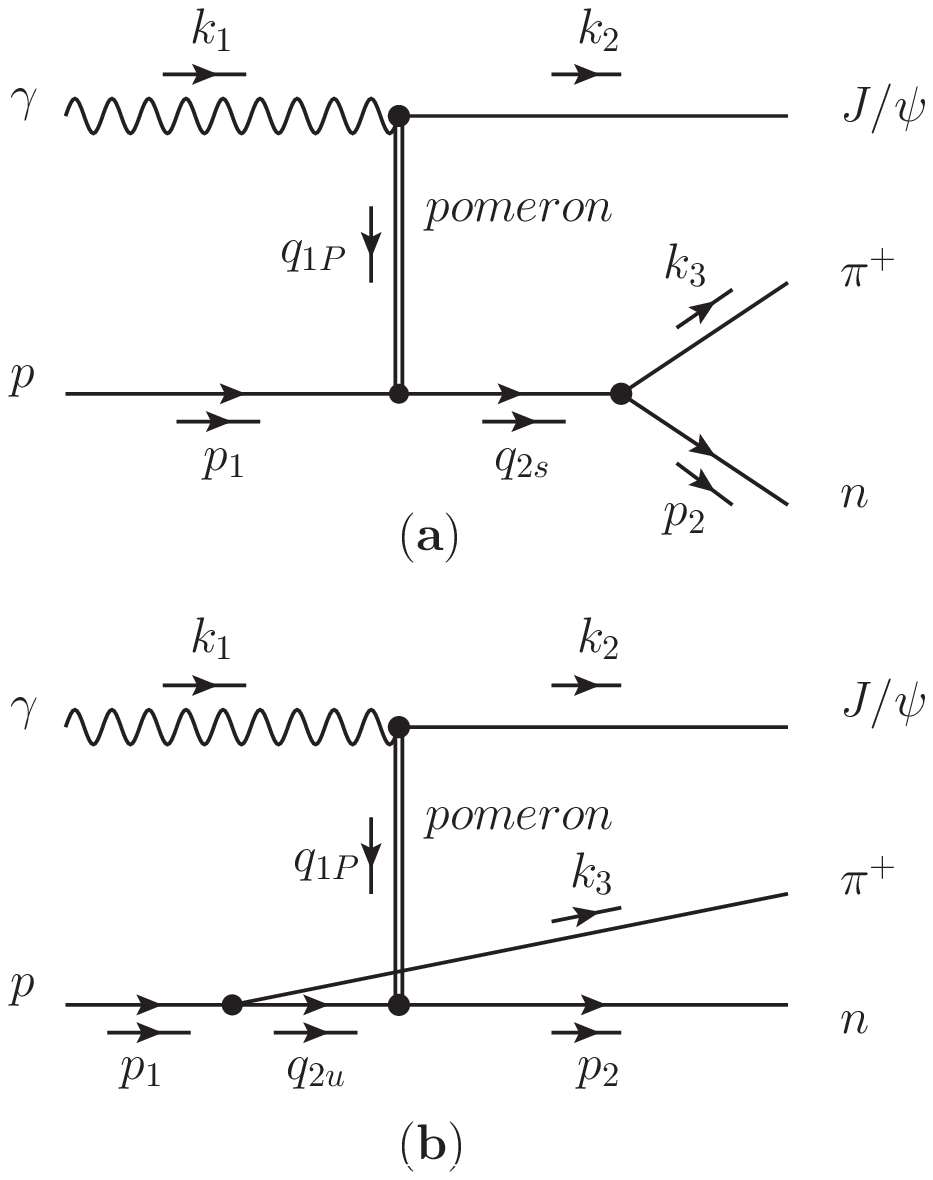}}
\caption{The $\gamma p\to J/\psi\pi^+n$ process through the Pomeron exchange.
\label{fig:2to3bg}}
\end{center}
\end{figure}

The Pomeron contribution has been widely applied in the study of the diffractive
transitions phenomenologically \cite{Donnachie:1987pu,Pichowsky:1996jx,Laget:1994ba}. Because the Pomeron can mediate the long-range interaction between a confined quark
and a nucleon, thus the Pomeron exchange can mainly contribute to the $\gamma p \to J/\psi p \to J/\psi\pi^+n$ reaction, where the Pomeron behaviors like an isoscalar photon with $C=+1$. The detailed description is shown in Fig. \ref{fig:2to3bg}.

To describe the Pomeron exchange process shown in \ref{fig:2to3bg}, we adopt the
formula given in Refs. \cite{Liu:2008qx,Donnachie:1987pu,Pichowsky:1996jx,Zhao:1999af}.
The Pomeron-nucleon coupling can be written as
\begin{equation}
F_\mu(t)=\frac{3\beta_0(4M_N^2-2.8t)}{(4M_N^2-t)(1-t/0.7)^2}\gamma_\mu= f(t)\gamma_\mu,
\end{equation}
where $t=q_{1P}^2=(k_1-k_2)^2$ denotes the the exchanged Pomeron momentum squared. $\beta_0$ is the
coupling constant between a single Pomeron and a light constituent
quark.

For the $\gamma V \mathcal{P}$ vertex, an
on-shell approximation for restoring the gauge invariance
is adopted as suggested in Ref.~\cite{Zhao:1999af}. Thus, the equivalent vertex for the
$\gamma\psi\mathcal{P}$ interaction is written as
\begin{eqnarray}
V_{\gamma V\cal{P}} &=& \frac{2\beta_c \times 4\mu_0^2}{(M_{\psi}^2
-t)(2\mu_0^2+M_{\psi}^2 -t)} T_{\mu\alpha\nu}\epsilon_\gamma^\mu\epsilon_\psi^{\nu}\cal{P}^\alpha  \nonumber \\
&=& V(t)T_{\mu\alpha\nu}\epsilon_\gamma^\mu\epsilon_\psi^{\nu}\cal{P}^\alpha,
\end{eqnarray}
where
\begin{eqnarray}
T^{\mu\alpha\nu} &=& (k_1+k_2)^\alpha g^{\mu\nu} -2k_1^\nu g^{\alpha\mu}+2\left[k_1^\mu g^{\alpha\nu}\right. \nonumber\\
&& \left. +\frac{k_2^\nu}{k_2^2}(k_1\cdot k_2 g^{\alpha\mu}-k_1^\alpha k_2^\mu-k_1^\mu k_2^\alpha)\right. \nonumber\\
&& \left. -\frac{k_1^2 k_2^\mu}{k_2^2 k_1\cdot k_2}(k_2^2 g^{\alpha\nu}-k_2^\alpha k_2^\nu)\right]+(k_1-k_2)^\alpha g^{\mu\nu}.
\end{eqnarray}
$\beta_c$ is the effective coupling constant between a Pomeron and
a charm quark within $J/\psi$, while $\mu_0$ is a cutoff in the form
factor related to the Pomeron.

And then the amplitudes of Fig.~\ref{fig:2to3bg}~(a) and (b) read as
\begin{eqnarray}
\mathcal{T}^{\mathcal{P}}_{afi} &=& \frac{g_{\pi NN}}{\sqrt{2}m_N} F_{pn\pi}(q_{2s}^2)f(t)V(t)\mathcal{G}_P(s,t)T^{\mu\alpha\nu}\nonumber\\
&& \times\bar{u}(p_2) \gamma_{5}\slashed{k}_3 \frac{\slashed{q}_{2s} + M_N}{q_{2s}^2 - M_N^2}
\gamma_{\alpha} u(p_1)\epsilon_{\gamma\mu}\epsilon_{\psi \nu}^\ast, \label{h1}\\
\mathcal{T}^{\mathcal{P}}_{bfi} &=& \frac{g_{\pi NN}}{\sqrt{2}m_N} F_{pn\pi}(q_{2u}^2)f(t)V(t)\mathcal{G}_P(s,t)T^{\mu\alpha\nu}\nonumber\\
&& \times\bar{u}(p_2) \gamma_{\alpha} \frac{\slashed{q}_{2u} + M_N}{q_{2u}^2 - M_N^2}
\gamma_{5}\slashed{k}_3 u(p_1)\epsilon_{\gamma\mu}\epsilon_{\psi \nu}^\ast,\label{h2}
\end{eqnarray}
where $\mathcal{G}_P(s,t)$ is related to the Pomeron trajectory
$\alpha(t)=1+\epsilon+\alpha^\prime t$, which satisfies
\begin{equation}
\mathcal{G}_P(s,t)= -i(\alpha^\prime s)^{\alpha(t)-1} \nonumber.
\end{equation}
The concrete values of the involved parameters in the above amplitudes include $\beta_0^2=4.0 \ \mbox{GeV}^2$, $\beta_c^2=0.8\ \mbox{GeV}^2$, $\alpha^\prime=0.25\ \mbox{GeV}^{-2}$, $\epsilon=0.08$, and $\mu_0=1.2\ \mbox{GeV}$.

Since the intermediate nucleon is off-shell just shown in Fig. \ref{fig:2to3bg}, a form factor for
the $pn\pi$ vertex is considered with the form
\begin{eqnarray}
F_{pn\pi}(q^2) &=&
\frac{\Lambda_{pn\pi}^2-M_N^2}{\Lambda_{pn\pi}^2 - q^2},
\end{eqnarray}
where $q^2$ is the momentum of the intermediate nucleon.
Later, we will discuss how to constrain the value of the
cutoff $\Lambda_{pn\pi}$.

With the amplitudes listed in Eq. (\ref{h0}) and Eqs. (\ref{h1})-(\ref{h2}), we obtain
the square of the total invariant transition amplitude
\begin{eqnarray}
|{\cal M}|^2 &=& \sum \left|T_{fi}^{Signal}+T_{afi}^\mathcal{P} +
T_{bfi}^\mathcal{P}\right|^2.
\end{eqnarray}
The corresponding total cross
section of the process $\gamma p\to J/\psi\pi^+n$ is
\begin{eqnarray}
 d\sigma &=& \frac{M_N^2}{|k_1\cdot p_1|}\frac{|\mathcal{M}|^2}{4}(2\pi)^4d\Phi_3(k_1+p_1;p_2,k_2,k_3)
\end{eqnarray}
with the definition of $n$-body phase space \cite{Beringer:1900zz}
$$d\Phi_n(P;p_1,...,p_n)=\delta^4(P-\sum\limits_{i=1}^np_i)\prod\limits_{i=1}^3\frac{d^3p_i}{(2\pi)^32E_i}.$$

\subsection{Numerical results}

\begin{figure}[htb]
\begin{center}
\scalebox{1.0}{\includegraphics[width=\columnwidth]{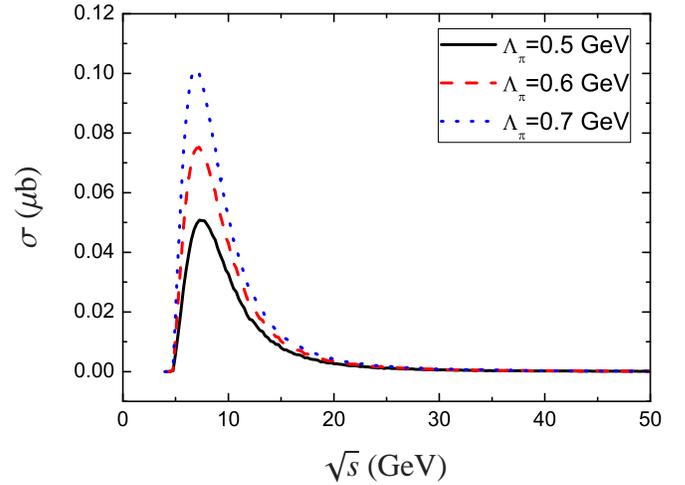}}
\caption{(color online). The cross sections of signal channel for $\gamma p \to J/\psi\pi^+n$ with different cutoff $\Lambda_\pi$. Here, the partial
width is taken to be $\Gamma[Z^+\to J/\psi\pi^+]=46$ MeV.
\label{fig:Res-cutoff}}
\end{center}
\end{figure}

\begin{figure}[htb]
\begin{center}
\scalebox{1.0}{\includegraphics[width=\columnwidth]{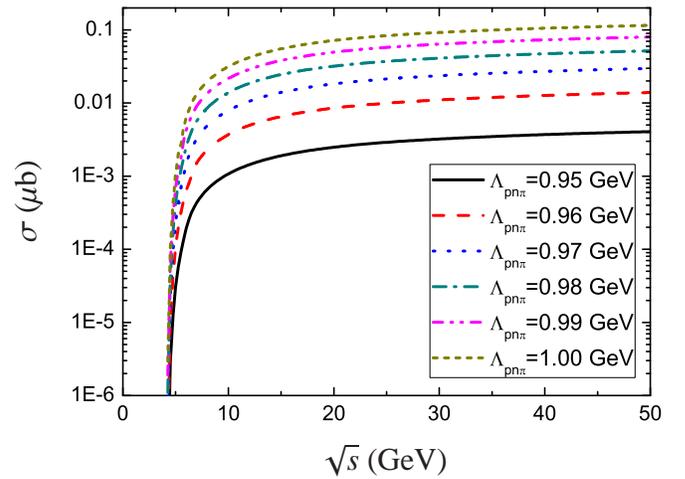}}
\caption{(color online). The contribution of background from the Pomeron exchange for $\gamma p \to J/\psi\pi^+n$ with different cutoff $\Lambda_{pn\pi}$.
\label{fig:Pom-cutoff}}
\end{center}
\end{figure}

\begin{figure}[htb]
\begin{center}
\scalebox{1.0}{\includegraphics[width=\columnwidth]{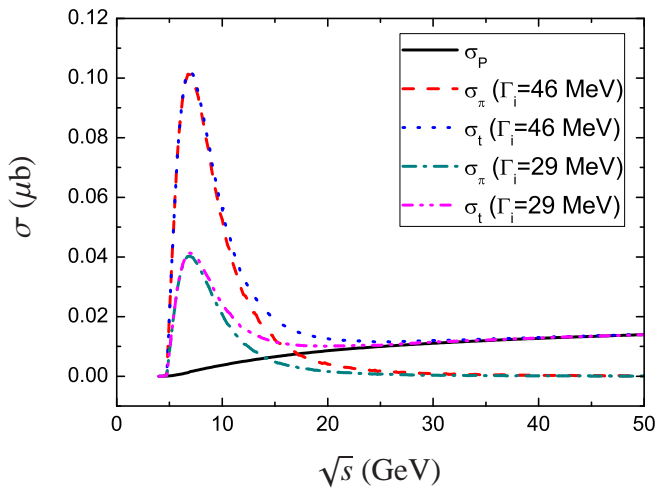}}
\caption{(color online). The energy dependence of the total cross sections for $\gamma p \to J/\psi\pi^+n$. Here, $\sigma_P$ and $\sigma_\pi$ are the results only considering the Pomeron exchange and the pion exchange contributions, respectively, while $\sigma_t$ denotes the total cross section of $\gamma p \to J/\psi\pi^+n$. We also give the variations of $\sigma_\pi$ and $\sigma_i$ to $\sqrt{s}$ corresponding to the typical values $\Gamma[Z^+\to J/\psi\pi^+]=29$ MeV and $\Gamma[Z^+\to J/\psi\pi^+]=46$ MeV.
\label{fig:2to3res}}
\end{center}
\end{figure}

With the help of the FOWL code and ROOT program \cite{root:5.34},
we present the obtained total cross section including both signal and background contributions
in Fig. \ref{fig:Res-cutoff}-\ref{fig:2to3res}.

First, we present the variation of the cross section from
the signal contribution for $\gamma p\to J/\psi\pi^+n$ with
different $\Lambda_{\pi}$ values as shown in Fig. \ref{fig:Res-cutoff},
where we take several typical values, i.e., $\Lambda_\pi=$ 0.5, 0.6, and 0.7 GeV.
The behavior of the result dependent on $\Lambda_\pi$ in Fig. \ref{fig:Res-cutoff} is similar to that in Fig. \ref{fig:2to2TCS-cutoff}.

Besides providing the information of signal channel, we also illustrate the Pomeron
exchange contributions with different cutoff $\Lambda_{pn\pi}$ as given in Fig. \ref{fig:Pom-cutoff},
where we take several typical values in the range of $\Lambda_{pn\pi}=0.95-1.0$ GeV with
step of 0.01 GeV. It is obvious that the cross sections are sensitive to the values of the cutoff.

We notice that there are experimental data for a cross section of very similar process $\gamma p\to J/\psi p$, where this cross section is just about 10 $nb$ \cite{JLAB,Levy:2007fb}.
If we naively expect that the process $\gamma p\to J/\psi \pi^+ n$ has similar or even lower cross section to that of $\gamma p\to J/\psi p$, $\Lambda_{pn\pi}$ can be roughly constrained, i.e., $\Lambda_{pn\pi}=0.96$ GeV. Thus, in the following discussion, we take typical values $\Lambda_\pi=$ 0.7 GeV and $\Lambda_{pn\pi}=$ 0.96 GeV. Then, we obtain the
total cross sections for $\gamma p\to J/\psi\pi^+n$, which are given in Fig. \ref{fig:2to3res}.
The line shape of the cross section of $\gamma p \to Z_c(3900)^+n \to J/\psi\pi^+n$
via the pion exchange (see Fig. \ref{fig:2to3res}) is rather similar to that of
$\gamma p \to Z_c(3900)^+n$ (see the bottom diagram in Fig. \ref{fig:2to2res}), which goes up very
rapidly near the threshold and has a peak around
$\sqrt{s}=7$ GeV. The line shape of the cross section for $\gamma p\to J/\psi\pi^+n$ via the
Pomeron exchange is monotonously increasing. There are obvious peaks appearing in the $\sqrt{s}$ dependence of the total cross section corresponding to both $\Gamma[Z^+\to J/\psi\pi^+]=29$ and 46 MeV, thus the signal
can be easily distinguished, which will be further illustrated later.
The overall cross sections corresponding to $\Gamma[Z^+\to J/\psi\pi^+]=29$ and 46 MeV are dominated by the $Z_c(3900)^+$ signal
at $\sqrt{s}\leq 14$ and 17 GeV, respectively, as shown in Fig. \ref{fig:2to3res}.

As a useful approach to the data analysis, the Dalitz plot can provide more abundant information compared with the former calculation of the total cross section. In the following, we perform the analysis of the Dalitz plot of the $\gamma p \to J/\psi\pi^+n$ process.

\begin{figure*}[htbp]
\begin{center}
\scalebox{1.0}{\includegraphics[width=\textwidth]{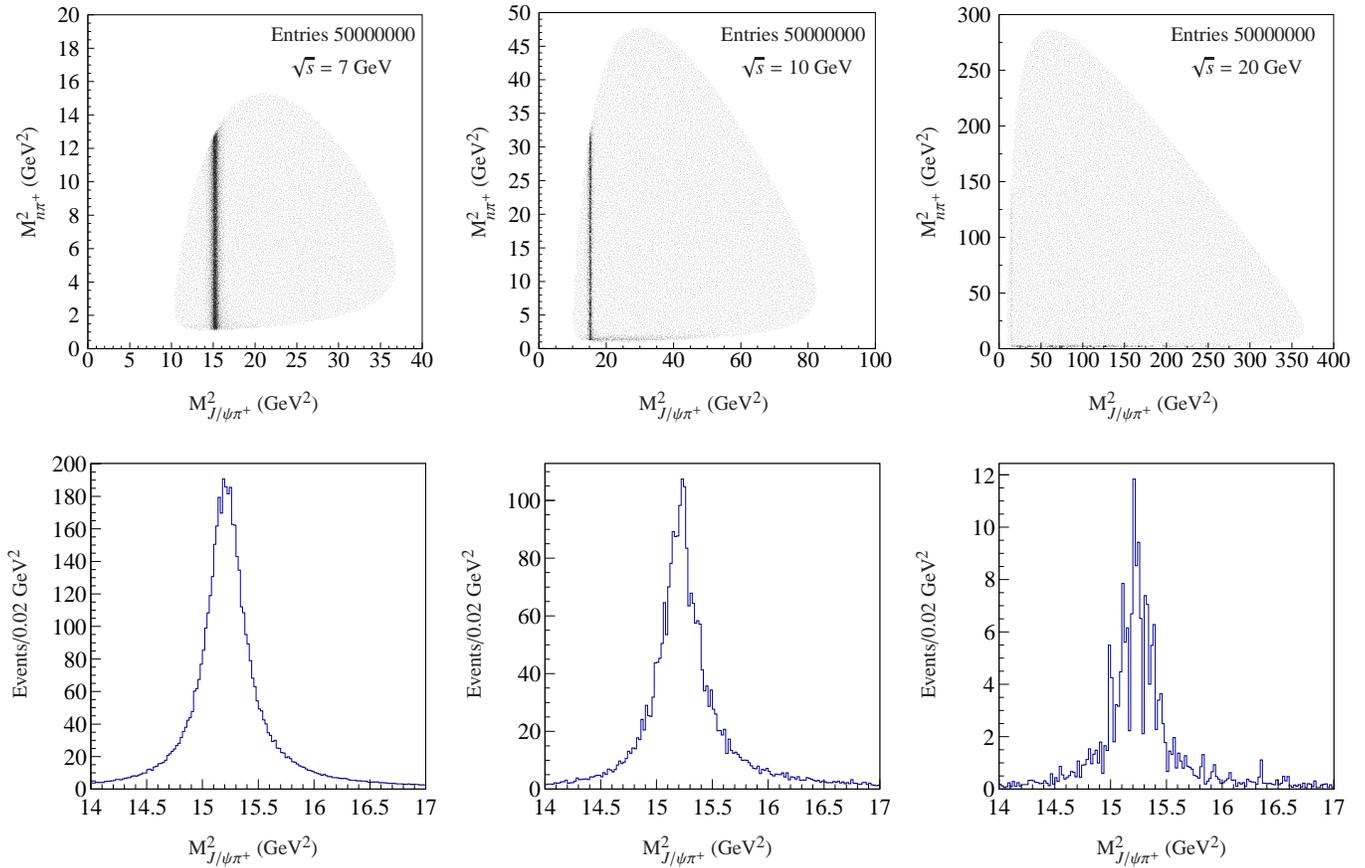}}
\caption{(color online). The Dalitz plot (top) and the corresponding $J/\psi\pi^+$ invariant mass spectrum (bottom)
for the $\gamma p \to J/\psi\pi^+n$ process. Here, all results correspond to the typical value
$\Gamma[Z^+\to J/\psi\pi^+]=29$ MeV.
\label{fig:Dalitz29}}
\end{center}
\end{figure*}

One can identify $Z_c(3900)^+$ by analyzing the $J/\psi\pi^+$ invariant mass spectrum from the Dalitz plot. With the results listed in Fig. \ref{fig:2to3res}, we find that the largest cross section for
$\gamma p\to Z_c(3900)^+n$ appears at $\sqrt{s}\simeq 7$ GeV. Thus, we take this energy as one of
the inputs to simulate the Dalitz plot of $\gamma p\to J/\psi\pi^+n$, where the $Z_c(3900)^+$ signal can
be distinguished in this kinematic region.
The Dalitz plot and the corresponding $J/\psi\pi^+$ invariant
mass spectrum with several typical values of $\sqrt{s}$ are presented in Fig. \ref{fig:Dalitz29},
where the partial decay width of $Z_c(3900)^+\to J/\psi\pi^+$ is taken as 29 MeV.
Figure \ref{fig:Dalitz29} indicates that there is an explicit vertical band corresponding to $Z_c(3900)^+$ when $\sqrt{s}=7$ GeV and $\sqrt{s}=10$
GeV. However, the band related to the $Z_c(3900)^+$ signal is reduced when $\sqrt{s}=20$ GeV, which is consistent with the result from Fig. \ref{fig:2to3res}. Here, the cross section $\sigma_\pi$ decreases with increasing $\sqrt{s}$ when $\sqrt{s}>7$ GeV. Thus, the $Z_c(3900)^+$ signal contribution can be ignored compared with the background contrition if $\sqrt{s}>20$ GeV. The Dalitz plot in Fig. \ref{fig:Dalitz29}
also shows the existence of the horizontal band with energies higher than 10 GeV, which is due to the Pomeron exchange
contribution, where the cross section from the Pomeron exchange goes up continuously.
By analyzing the $J/\psi\pi^+$ invariant mass spectrum, we also find that the number
of events of $Z_c(3900)^+$ can reach up to $200/0.02$ GeV$^2$ if taking 50 million collisions of $\gamma p$.

\begin{figure}[htb]
\begin{center}
\scalebox{1.0}{\includegraphics[width=\columnwidth]{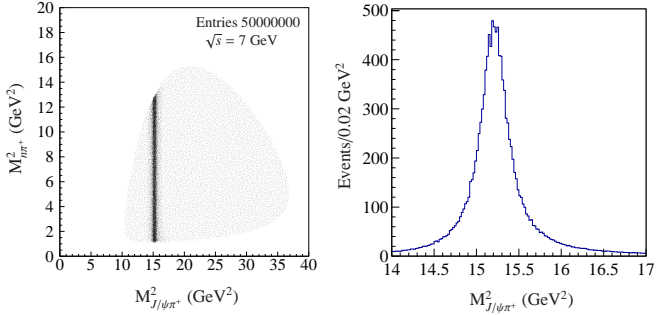}}
\caption{(color online). The Dalitz plot (left) and the $J/\psi\pi^+$ invariant mass spectrum (right)
for the $\gamma p \to J/\psi\pi^+n$ process when taking the typical value
$\Gamma[Z^+\to J/\psi\pi^+]=46$ MeV.
\label{fig:Dalitz46}}
\end{center}
\end{figure}

In Fig. \ref{fig:Dalitz46}, we also given the Dalitz plot and the $J/\psi\pi^+$ invariant mass spectrum taking
$\sqrt{s}=7$ GeV and $\Gamma[Z^+\to J/\psi\pi^+]=46$ MeV. The result shown in Fig. \ref{fig:Dalitz46} is similar to that in Fig. \ref{fig:Dalitz29}.
When taking 50 million collisions of $\gamma p$, the number of events of $Z_c(3900)^+$ can increase to 500/0.02 GeV$^2$.

\section{Discussion and conclusion}\label{sec:con}

Stimulated by the newly observed charged charmoniumlike state $Z_c(3900)^\pm$ \cite{Ablikim:2013mio,Liu:2013dau}, in this work we explore the discovery potential of the $Z_c(3900)^\pm$ production by meson photoproduciton, which is different from its $e^+e^-\to J/\psi\pi^+\pi^-$ process measured by BESIII and Belle \cite{Ablikim:2013mio,Liu:2013dau}.
We find that $\gamma p\to Z_c(3900)^+n$ can be as a suitable photoproduction process to study the production of $Z_c(3900)^+$, where the differential and total cross sections are calculated. One finds that the maximum of the total cross section of $\gamma p\to Z_c(3900)^+n$ is around the order of 0.1 $\mu b$, which is comparable with the cross section of $J/\psi$, $\psi(2S)$, and $\Upsilon$
photoproduction in the HERA experiment.

Furthermore, we carry out the analysis of the Dalitz plot of $\gamma p \to J/\psi\pi^+n$ and give the distribution of the corresponding $J/\psi\pi^+$ invariant mass spectrum, where the signal and background contributions are considered. The Dalitz plot analysis provides valuable information for further experimental study of $Z_c(3900)^+$ via meson photoproduction, where the $Z_c(3900)^+$ signal can be easily distinguished.

In summary, both the calculation of total reaction cross section and the analysis of the Dalitz plot indicate that it is accessible to search for $Z_c(3900)^+$ via meson photoproduction. We also expect further experimental progress on the  $Z_c(3900)^+$ photoproduction.

\section{Acknowledgements}
We would like to thank Dr. Alexey Guskov for his comments and communication of the experimental data of $\gamma p\to J/\psi p$ .
Q.Y.L also would like to thank Jun He, Ju-Jun Xie, Dian-Yong Chen and Xiao-Rui L\"{u}
for their valuable discussions and suggestions. This project is supported
by the National Natural Science Foundation of China under
Grants No. 11222547, No. 11175073, and No. 11035006,
the Ministry of Education of China (FANEDD under Grant
No. 200924, SRFDP under Grant No. 20120211110002,
NCET, the Fundamental Research Funds for the Central
Universities), and the Fok Ying-Tong Education
Foundation (No. 131006).

\vfill

\end{document}